%% file: 00paper.tex
\documentclass[sigconf]{acmart}

\usepackage{booktabs}

\usepackage[normalem]{ulem}  
\usepackage{bbold}  
\usepackage{subfigure}  
\usepackage{placeins}  
\usepackage{multirow} 
\usepackage{moreverb} 
\usepackage{enumerate} 
\usepackage{tabularx} 

\usepackage{tikz}
\usepackage{pgfplots}
\usetikzlibrary{arrows}
\usepackage{filecontents}
\usepackage{pgfplots,pgfplotstable}

\setcopyright{rightsretained}

\copyrightyear{2017}
\acmYear{2017}
\setcopyright{acmcopyright} 
\acmConference{ICTIR'17}{}{October 1--4, 2017, Amsterdam, Netherlands.} 
\acmPrice{15.00} 
\acmDOI{https://doi.org/10.1145/3121050.3121054}
\acmISBN{ISBN 978-1-4503-4490-6/17/10}

\begin{document}

\title{On Type-Aware Entity Retrieval}

\author{Dar\'{i}o Garigliotti}
\affiliation{University of Stavanger}
\email{dario.garigliotti@uis.no}

\author{Krisztian Balog}
\affiliation{University of Stavanger}
\email{krisztian.balog@uis.no}


\begin{abstract}
Today, the practice of returning entities from a knowledge base in response to search queries has become widespread.  
One of the distinctive characteristics of entities is that they are typed, i.e., assigned to some hierarchically organized type system (type taxonomy).
The primary objective of this paper is to gain a better understanding of how entity type information can be utilized in entity retrieval.
We perform this investigation in an idealized ``oracle'' setting, assuming that we know the distribution of target types of the relevant entities for a given query.
We perform a thorough analysis of three main aspects: (i) the choice of type taxonomy, (ii) the representation of hierarchical type information, and (iii) the combination of type-based and term-based similarity in the retrieval model.  
Using a standard entity search test collection based on DBpedia, we find that type information proves most useful when using large type taxonomies that provide very specific types.  We provide further insights on the extensional coverage of entities and on the utility of target types.  \end{abstract}



\maketitle

\input{ictir2017-types-01}  
\input{ictir2017-types-02}  
\input{ictir2017-types-03}  
\input{ictir2017-types-04}  
\input{ictir2017-types-05}  
\input{ictir2017-types-06}  
\input{ictir2017-types-07}  
\input{ictir2017-types-08}  
\input{ictir2017-types-09}  

\FloatBarrier  

\bibliographystyle{ACM-Reference-Format}
\bibliography{ictir2017-types}

\end{document}

%% file: ictir2017-types-01.tex
\section{Introduction}
\label{sec:intro}

Entities, such as people, organizations, or locations are natural units for organizing information; they can provide not only more focused responses, but often immediate answers, to many search queries~\cite{Pound:2010:AOR}.
Indeed, entities play a key role in transforming search engines into ``answer engines''~\cite{Mika:2013:ESW}.
The pivotal component that sparked this evolution is the increased availability of structured data published in knowledge bases, such as Wikipedia, DBpedia, or the Google Knowledge Graph, now primary sources of information for entity-oriented search. 
Major web search engines also shaped users' expectations about search applications; the single-search-box paradigm has become widespread, and ordinary users have little incentive (or knowledge) to formulate structured queries. 
The task we consider in this paper, 
referred to as \emph{ad-hoc entity retrieval}~\cite{Pound:2010:AOR}, corresponds to this setting: returning a ranked list of entities from a knowledge base in response to a keyword user query.

\begin{figure}[h]
	\centering
	\vspace{-0.1in}
	\includegraphics[width=0.37\textwidth]{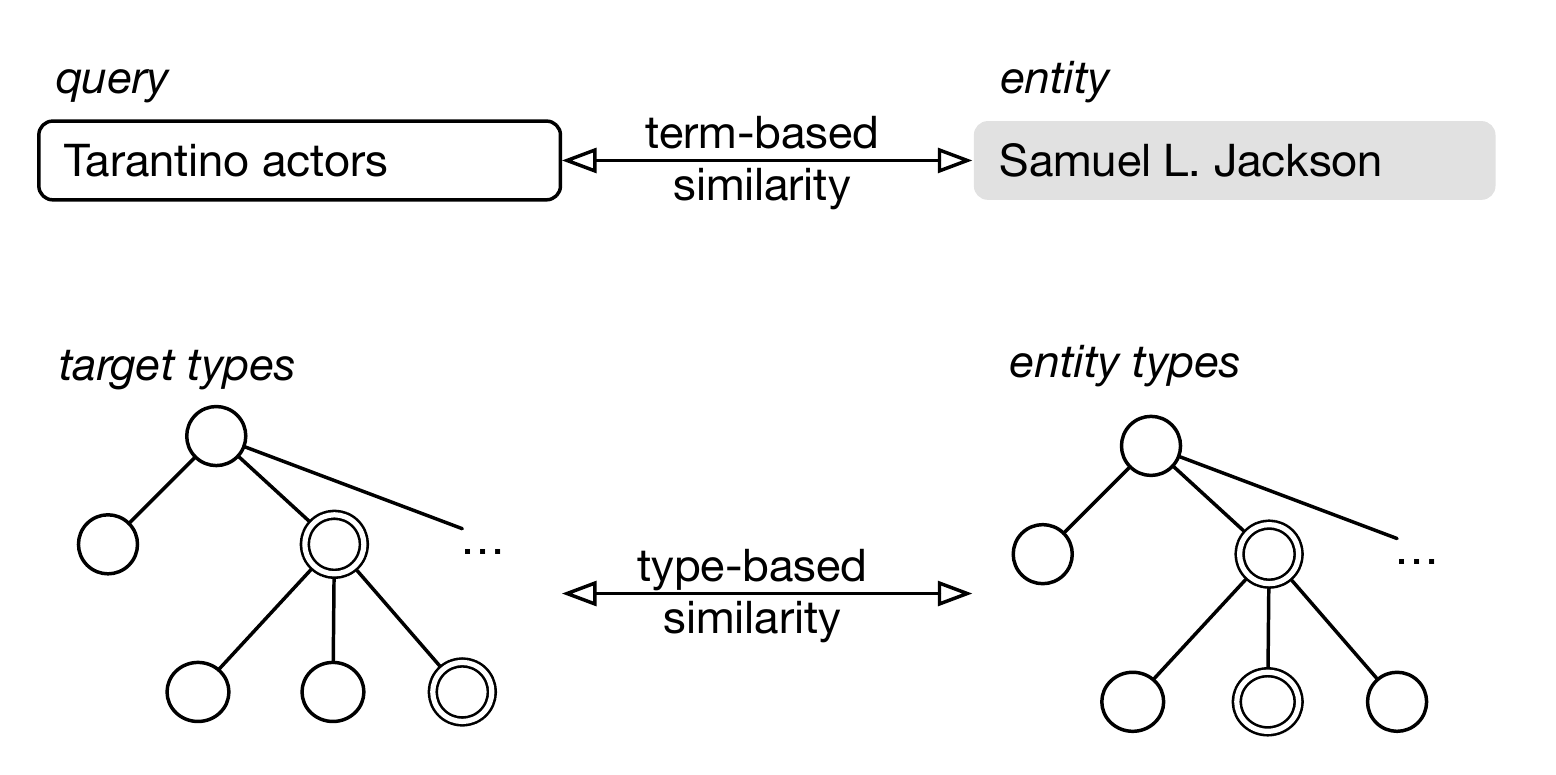}
	\caption{Entity retrieval using entity type information.}
	\label{fig:approach}
	\vspace{-0.05in}
\end{figure}

One of the unique characteristics of entity retrieval is that entities are typed, this is, grouped into more general classes, i.e., \emph{types}, of entities.
Types are typically organized in a hierarchy, which we will refer to as \emph{type taxonomy} hereinafter.
Each entity in the knowledge base can be associated with (i.e., is an \emph{instance of}) one or more types.
For example, in DBpedia, the type of the entity \texttt{Albert Einstein} is \texttt{Scientist}; according to Wikipedia, that entity belongs to the types \texttt{Theoretical physicists} and \texttt{People with acquired Swiss citizenship}, among others.  
It is assumed that by identifying the types of entities sought by the query (\emph{target types}, from now on), one can use this information to improve entity retrieval performance; see Figure~\ref{fig:approach} for an illustration. 
The main high-level research question we are concerned with in this study is: \emph{How to use entity type information in ad-hoc entity retrieval?}

\noindent
The concept of entity types, while seemingly straightforward, turns out to be a multifaceted research problem that has not yet been thoroughly investigated in the literature.  
Most of the research related with the usage of type information has been conducted in the context of the INEX Entity Ranking track~\cite{Demartini:2009:OIE}.  There, it is assumed that the user complements the keyword query with one or more target types, using Wikipedia's category system as the type taxonomy.  
The focus has been on expanding the set of target types based on hierarchical relationships and dealing with the imperfections of the type system~\cite{Demartini:2010:WFE,Balog:2011:QME,Pehcevski:2010:ERW,Kaptein:2013:ECS}.
Importantly, these developments have been motivated and driven by the peculiarities of Wikipedia's category system. It is not known whether the same methods prove effective, and even if these issues persist at all, in case of other type taxonomies.  One important contribution of this paper is that we consider and systematically compare multiple type taxonomies (DBpedia, Freebase, Wikipedia, and YAGO).
Additionally, there is the matter of representing type information, i.e., to what extent the hierarchy of the taxonomy should be preserved.  
Yet another piece of the puzzle is how to combine type-based and text-based similarity in the retrieval model.
Therefore, the research questions we address are as follows:

\begin{enumerate}[]  
	\item[] \textbf{RQ1} What is the impact of the particular choice of type taxonomy on entity retrieval performance?
	\item[] \textbf{RQ2} How to represent hierarchical entity type information for entity retrieval?
	\item[] \textbf{RQ3} How to combine term-based and type-based information?
\end{enumerate}
To answer the above questions, we conduct a series of experiments for all possible combinations of three dimensions: 
\begin{enumerate}[i)]
\itemsep 0pt
	\item The way term-based and type-based information is combined in the retrieval model; see Section~\ref{sec:dim_models}.
	\item The representation of hierarchical entity type information; see Section~\ref{sec:dim_reprs}.
	\item The choice of the type taxonomy; see Section~\ref{sec:dim_taxos}.
\end{enumerate}

\noindent
In summary, our work is the first comprehensive study on the usage of entity type information for entity retrieval.
Our main contributions are twofold. 
First, we present methods for representing types in a hierarchy, establishing type-based similarity, and combining term-based and type-based similarities.
Second, we perform a thorough experimental comparison and analysis of all possible configurations across the above identified three dimensions.
Our overall finding is that type information has the most benefits in case of large, deep type taxonomies that provide very specific types.  

%% file: ictir2017-types-02.tex
\section{Related Work}
\label{sec:rel}

The task of entity ranking has been studied in different flavors, including ad-hoc entity retrieval~\citep{Pound:2010:AOR,Neumayer:2012:MEA}, list search~\citep{Balog:2012:OTE,Demartini:2010:OIE}, related entity finding~\citep{Balog:2010:OTE}, and question answering~\citep{Lopez:2013:EQA}.
Our interest in this work lies in the usage of type information for general-purpose entity retrieval against a knowledge base (KB), where queries may belong to either of the above categories. 

\label{subsec:rel_er}

\paragraph{Retrieval models.} Early works represented type information as a separate field in a fielded entity model~\citep{Zhu:2008:IDF}.  
In later works, types are typically incorporated into the retrieval method by combining term-based similarity with a separate type-based similarity component.  This combination may be done using (i) a linear interpolation~\citep{Balog:2011:QME,Kaptein:2013:ECS,Pehcevski:2010:ERW} or (ii) in a multiplicative manner, where the type-based component essentially serves as a filter~\citep{Bron:2010:RRE}.
\citet{Raviv:2012:RFE} introduce a particular version of interpolation using Markov Random Fields, linearly aggregating each of the scores for the joint distribution of the query with entity document, type, and name.
All the mentioned works have consistently reported significant performance improvements when a type-based component is incorporated into the (term-based) retrieval model.  However, type-aware approaches have not been systematically compared to date.
We formalize these two general combination strategies, interpolation and filtering, in Section~\ref{sec:dim_models}, and then compare them in Section~\ref{sec:results}.

\paragraph{Type taxonomies.} The choice of a particular type taxonomy is mainly motivated by the problem setting, depending on whether a wide-coverage type system (like Wikipedia categories) or a curated, well-designed ontology (e.g., the DBpedia Ontology) is desired.  The most common type system used in prior work is Wikipedia categories~\cite{Demartini:2010:WFE,Balog:2011:QME,Kaptein:2013:ECS,Raviv:2012:RFE,Bron:2010:RRE}.  This is in part for historical reasons, as this was the underlying type system used at the INEX Entity Ranking track, where type information was first exploited.  Further choices include the DBpedia Ontology~\cite{Balog:2012:HTT,Tonon:2013:TRE}, YAGO types~\cite{Demartini:2010:WFE,Sawant:2013:LJQ,Tonon:2013:TRE,Nakashole:2013:FST}, Freebase~\cite{Lin:2012:NNP}, and schema.org~\cite{Tonon:2013:TRE}.  We are not aware of any work that compared different type taxonomies for entity retrieval.

\paragraph{Representations of type information.} Target types are commonly considered either as a set~\cite{Pehcevski:2010:ERW, Demartini:2010:WFE, Raviv:2012:RFE, Kaptein:2013:ECS} or as a bag (weighted set)~\cite{Vallet:2008:IMI,Balog:2011:QME, Sawant:2013:LJQ}. 
Various ways of measuring type-based similarity have been proposed~\citep{Vercoustre:2008:UWC,Kaptein:2009:FEW,Weerkamp:2009:AGL, Zhu:2008:IDF, Demartini:2008:LIQ}. In this work we employ a state-of-the-art probabilistic approach by~\citet{Balog:2011:QME} (cf. Section~\ref{sec:dim_models:type_based}).
Within a taxonomy, types are arranged in a hierarchy.    
Several approaches have attempted to expand the set of target types based on the hierarchical structure of the type system~\cite{Pehcevski:2010:ERW,Balog:2011:QME,Bron:2010:RRE, Demartini:2010:WFE}.
Importantly, the investigation of type hierarchies has been limited to Wikipedia, and, even there, mixed results are reported~\cite{Vercoustre:2008:UWC,Zhu:2008:IDF,Demartini:2008:LIQ,Jamsen:2008:ERB}. 
It remains an open question whether considering the hierarchical nature of types benefits retrieval performance. We aim to fill that gap.

\label{subsec:rel_tti}

\paragraph{Target Type Identification}
The INEX Entity Ranking track~\citep{Demartini:2010:OIE} and the TREC Entity track~\citep{Balog:2012:OTE} both featured scenarios where target types are provided by the user.  
In the lack of explicit target type information, one might attempt to infer types from the keyword query.  This subtask is introduced by~\citet{Vallet:2008:IMI} as the \emph{entity type ranking} problem.  They extract entity mentions from the set of top relevant passages, then consider the types associated with the top-ranked entities using various weighting functions. \citet{Kaptein:2010:ERU} similarly use a simple entity-centric model. 
Manually assigned target types tend to be more general than automatically identified ones~\cite{Kaptein:2013:ECS}. 
Having a hierarchical structure, therefore, makes it convenient to assign more general types.  
In \cite{Balog:2012:HTT}, a hierarchical version of the \emph{target type identification} task is addressed using the DBpedia Ontology and language modeling techniques.
\citet{Sawant:2013:LJQ} focus on telegraphic queries and assume that each query term is either a type hint or a ``word matcher.''  They consider multiple interpretations of the query and tightly integrate type detection within the ranking of entities.  Their approach further relies on the presence of a large-scale web corpus.
In our case, an oracle process generates the query target type distribution from its set of known relevant entities (cf. Section~\ref{sec:expsetup:oracle}).

\paragraph{Entity Types}
\label{subsec:rel_et}

A further complicating issue is that the type information associated with entities in the knowledge base is incomplete, imperfect, or missing altogether for some entities.  
Automatic typing of entities is a possible solution for alleviating some of these problems.
For example, approaches to extend entity type assignments in DBpedia include mining associated Wikipedia articles for wikilink relations~\citep{Nuzzolese:2012:TIT}, patterns over logical interpretations of the deeply parsed natural language definitions~\citep{Gangemi:2012:ATD}, or linguistic hypotheses about category classes~\citep{Fossati:2015:ULE}. 
Several works have addressed entity typing over progressively larger taxonomies with finer-grained types~\citep{Fleischman:2002:FGC, Giuliano:2009:FCN, Rahman:2010:IFS, Ling:2012:FER,Yosef:2012:HHT}. 
Regarding the task of detecting and typing \emph{emerging entities}, having fine-grained types for new entities is of particular importance for informative knowledge~\cite{Lin:2012:NNP,Nakashole:2013:FST}.  
   

%% file: ictir2017-types-03.tex
\section{Type-aware Entity Retrieval}
\label{sec:dim_models}  

In this section we formally describe the type-aware entity retrieval models we will be using for investigating the research questions stated in Section \ref{sec:intro}.  Our contributions do not lie in this part; the techniques we present were shown to be effective in prior research.

We formulate our retrieval task in a generative probabilistic framework.  Given an input query $q$, we rank entities $e$ according to

\begin{equation}
   P(e|q) \propto P(q|e)P(e) ~. \label{eq:Peq}
\end{equation}
When uniform entity priors are assumed, the final ranking of entities boils down to the estimation of $P(q|e)$.
We consider the query in the term space as well as in the type space.  Hence, we write $q=(q_w,q_t)$, where $q_w$ holds the query terms (words) and $q_t$ holds the \emph{target types}. 
Two ways of factoring the probability $P(q|e)$ are presented in Section~\ref{sec:dim_models:models}.  All models share two components: term-based similarity, $P(q_w|e)$, and type-based similarity, $P(q_t|e)$.  These are discussed in Sections~\ref{sec:dim_models:term_based} and~\ref{sec:dim_models:type_based}, respectively.

\subsection{Retrieval Models}
\label{sec:dim_models:models}

We present two alternative approaches for combining term-based and type-based similarity.

\subsubsection{Filtering}

Assuming conditional independence between the term-based and type-based components, the final score becomes a multiplication of the components:
\begin{equation}
   P(q|e) = P(q_w|e) P(q_t|e) ~.
\end{equation}
This approach is a generalization, among others, of the one used in \cite{Bron:2010:RRE} (where the term-based information itself is unfolded into multiple components, considering not only language models from textual context but also estimations of entity co-occurrences).
We consider two specific instantiations of this model: 

\begin{description}
 	\item[Strict filtering] where $P(q_t|e)$ is $1$ if the target types and entity types have a non-empty intersection, and is $0$ otherwise.
 	\item[Soft filtering] where $P(q_t|e) \in [0..1]$ and is estimated using the approach detailed below in~Section~\ref{sec:dim_models:type_based}.
\end{description}

\subsubsection{Interpolation}

Alternatively, a mixture model may be used, which allows for controlling the importance of each component.  Nevertheless, the conditional independence between $q_w$ and $q_t$ is still imposed by this model:
\begin{equation}
   P(q|e) = (1-\lambda_t) P(q_w|e) + \lambda_t P(q_t|e) ~.
\end{equation}
Examples of this approach include \cite{Pehcevski:2010:ERW,Balog:2011:QME,Raviv:2012:RFE,Kaptein:2013:ECS}.

\subsection{Term-based Similarity}
\label{sec:dim_models:term_based}

We base the estimation of the term-based component, $P(q_w|e)$, on statistical language modeling techniques since they have shown to be an effective approach in prior work, see, e.g.,~\cite{Balog:2011:QME,Kaptein:2013:ECS,Bron:2010:RRE,Balog:2013:TCE}. 
Specifically, we employ the Mixture of Language Models method from~\cite{Balog:2013:TCE} with two fields, title and content. Following \cite{Neumayer:2012:SGE}, the weights are set to $0.2$ and $0.8$, respectively. This is a simple, yet solid baseline approach.
We note that the term-based component is not the focus of this work; any other approach could also be plugged in (provided that the retrieval scores are mapped to probabilities).

\subsection{Type-based Similarity}
\label{sec:dim_models:type_based}

Rather than considering types simply as a set, we assume a distributional representation of types, also referred to as \emph{bag-of-types}.  Namely, a type in the bag may occur with repetitions, naturally rendering it more important.
Following~\cite{Balog:2011:QME}, we represent type information as a multinomial probability distribution over types, both for queries and for entities.  Specifically, let $\theta_q$ denote the target type distribution for the query $q$ (such that $\sum_t P(t|\theta_q)=1$).  We assume that there is some mechanism in place that estimates this distribution; in our experiments, we will rely on an ``oracle'' that provides us exactly with this information (cf. Section~\ref{sec:expsetup:oracle}).  Further, let $\theta_e$ denote the target type distribution for entity $e$.  We assume that a function $n(t,e)$ is provided, which returns $1$ if $e$ is assigned to type $t$, otherwise $0$. We present various ways of setting $n(t,e)$ based on the hierarchy of the type taxonomy in Section~\ref{sec:dim_reprs}.  We note that $n(t,e)$ is not limited to having a binary value; this quantity could, for example, be used to reflect how important type $t$ is for the given entity $e$.  We use a multinomial distribution to allow for such future extensions.  Based on these raw counts, the type-based representation of an entity $e$ is estimated using Dirichlet smoothing:
\begin{equation}\label{eq:ptthetae}
   P(t|\theta_e) = \frac{n(t,e) + \mu P(t)}{\sum_{t'}{n(t',e)} + \mu} ~,
\end{equation}
where $P(t)$ is the background type model obtained by a maximum-likelihood estimate, and $\mu$ is the smoothing parameter, which we set to the average number of types assigned to an entity.

With both $\theta_q$ and $\theta_e$ in place, we estimate type-based similarity using the Kullback-Leibler (KL) divergence of the two distributions:
\begin{equation}
   P(q_t|e) = z (\max_{e'} KL(\theta_q\|\ \theta_{e'}) - KL(\theta_q\|\theta_e)) ~,
   \label{eq:typesim}
\end{equation}
where $z$ is a normalization factor.  Note that the smaller the divergence the more similar the distributions are, therefore in Eq.~\eqref{eq:typesim} we subtract it from the maximum KL-divergence, in order to obtain a probability distribution.  For further details we refer to \cite{Balog:2011:QME}.

%% file: ictir2017-types-04.tex
\section{Representing Hierarchical Entity Type Information}
\label{sec:dim_reprs}

\begin{figure*}[!th]
    \centering
	\begin{tabular}{ c c c }
		\subfigure[Path-to-top type(s)]{
	        \includegraphics[width=0.23\textwidth]{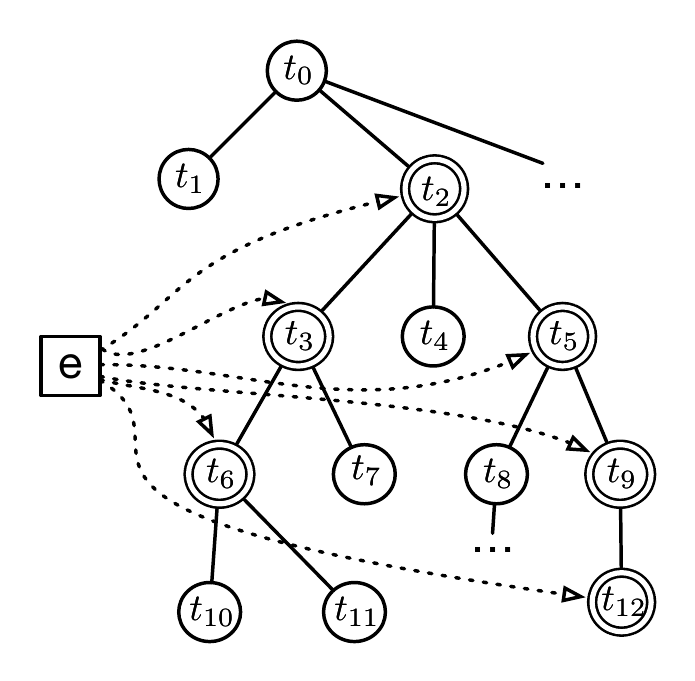}
			\label{fig:reprs:mode_path}
		}
		&
		\hspace{0.2in}  
		\subfigure[Top-level type(s)]{
	        \includegraphics[width=0.23\textwidth]{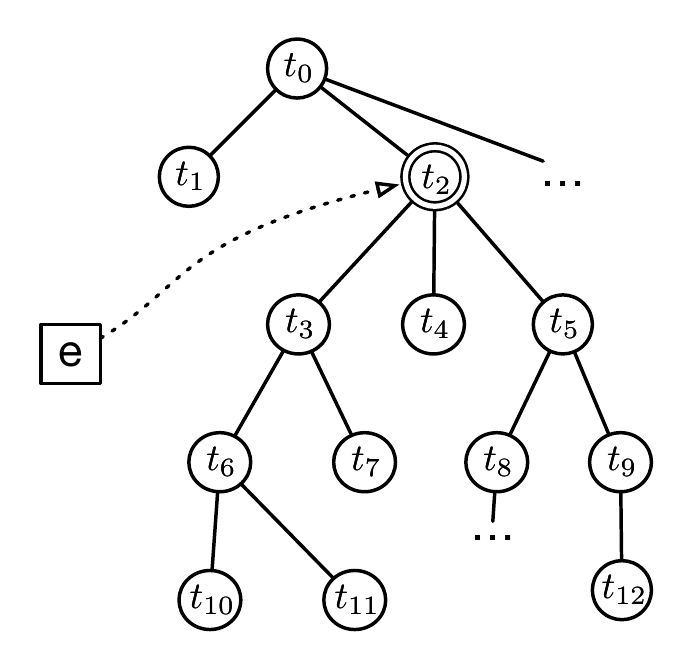}
			\label{fig:reprs:mode_top}
		}
		&
		\hspace{0.2in}  
		\subfigure[Most specific type(s)]{
	        \includegraphics[width=0.23\textwidth]{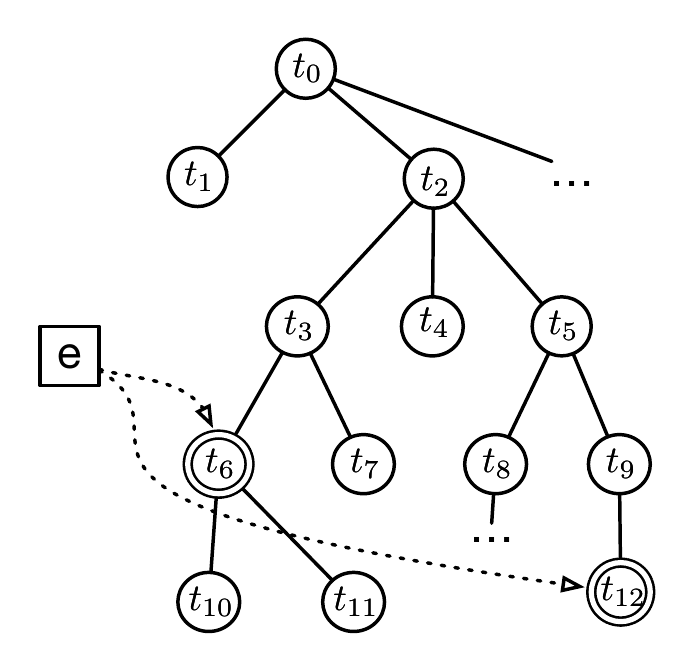}
			\label{fig:reprs:mode_most}
		}	
	\end{tabular}
	\centering
	\vspace*{-0.5\baselineskip}
	\caption{Alternative ways of representing entity-type assignments with respect to the type taxonomy. The dashed arrows point to the types that are assigned to entity $e$. The root node of the  taxonomy is labeled with $t_0$.}
	\label{fig:modes}
\end{figure*}

This section presents various ways of representing hierarchical entity type information.  That is, how to set the quantity $n(t,e)$, which is needed for estimating type-based similarity between target types of the query and types assigned to the entity in the knowledge base.
Before proceeding further, let us introduce some terminology and notation.  
\begin{itemize}
\itemsep 0pt
	\item $T$ is a type taxonomy that consists of a set of hierarchically organized entity types, and $t \in T$ is a specific entity type.
	\item $E$ is the set of all entities in the knowledge base, and $e \in E$ is a specific entity.
	\item $T(e)$ is the set of types that are assigned to entity $e$ in the knowledge base. We refer to this as a set of \emph{assigned types}. Note that $T(e)$ might be an empty set. 
\end{itemize}
We impose the following constraints on the type taxonomy.
\begin{enumerate}[i)]
\itemsep 0pt
	\item There is a single root node $t_0$ that is the ancestor of all types (e.g., \texttt{<owl:Thing>}). Since all entities belong to this type, it is excluded from the set of assigned types by definition.
	\item We restrict the type taxonomy to subtype-supertype relations; each type $t$ has a single parent type denoted as $\pi(t)$.
	\item Type assignments are transitive, i.e., an entity that belongs to a given type also belongs to all ancestors of that type: $t \in T(e) \wedge \pi(t) \neq t_0 \implies \pi(t) \in T(e)$.
\end{enumerate}
We further note that an entity might belong to multiple types under different branches of the taxonomy. Assume that $t_i$ and $t_j$ are both types of $e$. It might be then that their nearest common ancestor in the type hierarchy is $t_0$.

While $T(e)$ holds the types assigned to entity $e$, there are multiple ways of turning it into a numerical value, $n(t,e)$, which reflects the type's importance or weight with respect to the given entity.  This weight is taken into account when building the type-based entity representation in Eq.~\eqref{eq:ptthetae}.  In this work, we treat all types equally important for an entity, i.e., use binary values for $n(t,e)$.

We consider the following three options for representing hierarchical type information; see Figure~\ref{fig:modes} for an illustration.  
In our definitions, we use $\mathbb{1}(x)$ as an indicator function, which returns the value $1$ if condition $x$ is true and returns $0$ otherwise.
\begin{description}
	\item [Path-to-top] It counts all types that are assigned to the entity in the knowledge base, excluding the root (from constraint (iii) it follows that $T(e)$ contains all the types to the top-level node):
$$n(t,e) = \mathbb{1}\big(t \in T(e)\big) ~.$$

	\item [Top-level type(s)] Only top-level types are considered for an entity, that is, types that have the root node as their parent:
$$n(t,e) = \mathbb{1}\big( t \in T(e) \wedge \pi(t)=t_0 \big) ~.$$

	\item [Most specific type(s)] From each path, only the most specific type is considered for the entity:
$$n(t,e) = \mathbb{1}\big(t \in T(e) \wedge \nexists t' \in T(e): \pi(t')=t\big) ~.$$

\end{description}
Even though there may be alternative representations, these three are natural ways of encoding hierarchical information.

%% file: ictir2017-types-05.tex
\section{Entity Type Taxonomies}
\label{sec:dim_taxos}  

In this paper we study multiple type taxonomies from various knowledge bases: DBpedia, Freebase, Wikipedia, and YAGO.
These vary a lot in terms of hierarchical structure and in how entity-type assignments are recorded.
We normalize these type taxonomies to a uniform structure, adhering to the constraints specified in Section~\ref{sec:dim_reprs}.
Table~\ref{table:type_taxonomies_stats} presents an overview of the type systems (after normalization).   
The number of type assignments are counted according to the path-to-top representation.  Properties of the four type systems and details of the normalization process are discussed below.

\begin{table}[t]
  \centering
  \caption{Overview of normalized type taxonomies and their statistics. The top block is about the taxonomy itself; the bottom block is about type assignments of entities.}
  \label{table:type_taxonomies_stats}
  \begin{tabular}{l | r r r r}
    \toprule
    Type system & DBpedia & Freebase & \vtop{\hbox{\strut Wikipedia}\hbox{\strut categories}} & YAGO\\
    \midrule
	\#types & 
		713 & 1,719 & 423,636 & 568,672 \\
	\#top-level types & 
		22 & 92 & 27 & 61 \\
	\#leaf-level types & 
		561 & 1,626 & 303,956 & 549,754 \\
	height & 
		7 & 2 & 35 & 19 \\
	\midrule
	\#types used & 
		408 & 1,626 & 359,159 & 314,632 \\	
	\#entities w/ type & 
		4.87M & 3.27M & 3.52M & 2.88M \\
	avg \#types/entity & 
		2.8 & 4.4 & 20.8 & 13.4 \\
	mode depth & 
		2 & 2 & 11 & 4 \\
    \bottomrule
   \end{tabular}  
\end{table}

\subsection{Type Taxonomies}

\paragraph*{Wikipedia categories}
The Wikipedia category system, developed and extended by Wikipedia editors, consists of textual labels known as categories. This categorization is not a well-defined ``is-a'' hierarchy, but a graph; a category may have multiple parent categories and there might be cycles along the path to ancestors~\cite{Kaptein:2010:ERU}.  Also, categories often represent only loose relatedness between articles; category assignments are neither consistent nor complete~\cite{Demartini:2009:OIE}.

We transformed the Wikipedia category graph, consisting of over 1.16M categories, into a type taxonomy as follows.  First, we selected a set of 27 top-level categories covering most of the knowledge domains.\footnote{The selected top-level categories are the main categories for each section of the portal \url{https://en.wikipedia.org/wiki/Portal:Contents/Categories}.  (As an alternative, we also considered the categories from \url{https://en.wikipedia.org/wiki/Category:Main_topic_classifications}, and found that it comprises a similar category selection).} These became the top-level nodes of the taxonomy, all with a single common root type \texttt{<owl:Thing>}.  All super-categories that these selected top-level categories might have in the graph were discarded.  Second, we removed multiple inheritances by selecting a single parent per category.  For this, we considered the population of a category to be the set of its assigned articles.  Each category was linked in the taxonomy with a single parent in the graph whose intersection between their populations is the maximal among all possible parents; in case of a tie, the most populated parent was chosen.  Under this criterion, and for the purpose of understanding hierarchical relations, any category without a parent was discarded.  Lastly, from this partial hierarchy (which is still a graph, not a tree), we obtained the final taxonomy by performing a depth-first exploration from each top-level category, and avoiding to add those arcs that would introduce cycles. The resulting taxonomy contains over 423K categories and reaches a maximum depth of 35 levels.\footnote{We have confirmed experimentally that enforcing the Wikipedia category graph to satisfy the taxonomical constraints does not hurt retrieval performance. In fact, it is the opposite: it results in small, but statistically significant improvements.}

\paragraph*{DBpedia ontology}
The DBpedia Ontology is a well-designed hierarchy since its inception; it was created manually by considering the most frequently used infoboxes in Wikipedia.  It continues to be properly curated to address some weaknesses of the Wikipedia infobox space.  While the DBpedia Ontology is clean and consistent, its coverage is limited to entities that have an associated infobox.  It consists of 712 classes organized in a hierarchy of 7 levels.  

\paragraph*{YAGO taxonomy}
YAGO is a huge semantic knowledge base, derived from Wikipedia, WordNet, and GeoNames~\cite{Suchanek:2007:YCS}.  Its classification schema is constructed by taking leaf categories from the category system of Wikipedia and then using WordNet synsets to establish the hierarchy of classes. The result is a deep subsumption hierarchy, consisting of over 568K classes.
We work with the YAGO taxonomy from the current version of the ontology (3.0.2).  We normalized it by adding a root node, \texttt{<owl:Thing>}, as a parent to every top-level type. 
           
\paragraph*{Freebase types}
Freebase has a two-layer categorization system, where types on the bottom level are grouped under high-level domains.  
We used the latest public Freebase dump (2015-03-31), discarding domains meant for administering the Freebase service itself (e.g.; \texttt{base}, \texttt{common}).  Additionally, we made \texttt{<owl:Thing>} the common root of all the domains, and finally obtained a taxonomy of 1,719 types.

\subsection{Entity-Type Assignments}

Now that we have presented the four type taxonomies, we also need to discuss how type assignments of entities are obtained.
We use DBpedia 2015-10 as our knowledge base, which makes DBpedia types, Wikipedia categories, and YAGO type assignments readily available. 
For the fourth type taxonomy, Freebase, we followed same-as links from DBpedia to Freebase (which exist for 95\% of the entities in DBpedia) and extracted type assignments from Freebase. 
It should be noted that entity-type assignments are provided differently for each of these taxonomies; DBpedia and Freebase supply a single (most specific) instance type for an entity, Wikipedia assignments include multiple categories for a given entity (without any restriction), while YAGO adheres to the path-to-top representation.   
We treat all entity-type assignments transitively, adhering to constraint (iii) in Section~\ref{sec:dim_reprs}.

%% file: ictir2017-types-06.tex
\section{Experimental Setup}
\label{sec:expsetup}

We base our experiments on the DBpedia knowledge base (version 2015-10).
DBpedia~\cite{Lehmann:2015:DAL}, as a central hub in the Linked Open Data cloud, provides a large repository of entities, which are mapped---directly or indirectly---to each of the type taxonomies of interest.  

\label{sec:expsetup:collection}

\paragraph{Test Collection.}
Our experimental platform is based on the test collection developed in~\cite{Balog:2013:TCE}. The dataset contains 485 queries, synthesized from various entity-related benchmarking evaluation campaigns. These range from short keyword queries to natural language questions. 

\label{sec:expsetup:oracle}

\paragraph{Target Types Oracle.}
Throughout all our experiments, we make use of a so-called \emph{target type oracle}.  We assume that there is an ``oracle'' process in place that provides us with the (distribution of) correct target types for a given query.  This corresponds to the setting that was employed at previous benchmarking campaigns (such as the INEX Entity Ranking track~\cite{Demartini:2009:OIE} and the TREC Entity track~\cite{Balog:2012:OTE}), where target types are provided explicitly as part of the topic definition.
We need this idealized setting to ensure that our results reflect the full potential of using type information, without being hindered by the imperfections of an automated type detector.

For a given query $q$, we take the union of all types of all entities that are judged relevant for that query.  Each of these types $t$ becomes a target type, and its probability $P(t|\theta_q)$ is set proportional to the number of relevant entities that have that type.

\label{sec:expsetup:comparison}

\paragraph{Retrieval Models.}
As our baseline, we use a term-based approach, specifically the Mixture of Language Models~\cite{Balog:2013:TCE}, which we described in Section~\ref{sec:dim_models:term_based}.
We compare three type-aware retrieval models (cf. Section~\ref{sec:dim_models:models}): strict filtering, soft filtering, and interpolation.  For the latter, we perform a sweep over the possible type weights $\lambda_t \in [0, 1]$ in steps of 0.05, and use the best performing setting when comparing against other approaches.  (Automatically estimating the $\lambda_t$ parameter is outside the scope of this work.)

\label{sec:expsetup:allVS4tt}

\paragraph{Type Assignments.}
To ensure that the differences we observe are not a result of missing type assignments, we distinguish between two settings in our experiments. 

\begin{description}
	\item [4TT] We restrict our set of entities to those that have types assigned to them from all four type systems (1.51M entities in total).  This ensures that the results we obtain are comparable across the different type systems. 
We also restrict the set of queries to those that have target types in all four type systems; queries without any relevant results (as a consequence of these restrictions) are filtered out.  This leaves us with a total of 419 queries.
	\item [ALL] We include all entities from the knowledge base and use the original set of relevance assessments without any modifications.  Hence, some entities and queries do not have types assigned from one or more taxonomies.    
\end{description}

%% file: ictir2017-types-07.tex
\section{Results}
\label{sec:results}

\begin{figure*}[!th]
    \centering
    \includegraphics[width=0.9\textwidth]{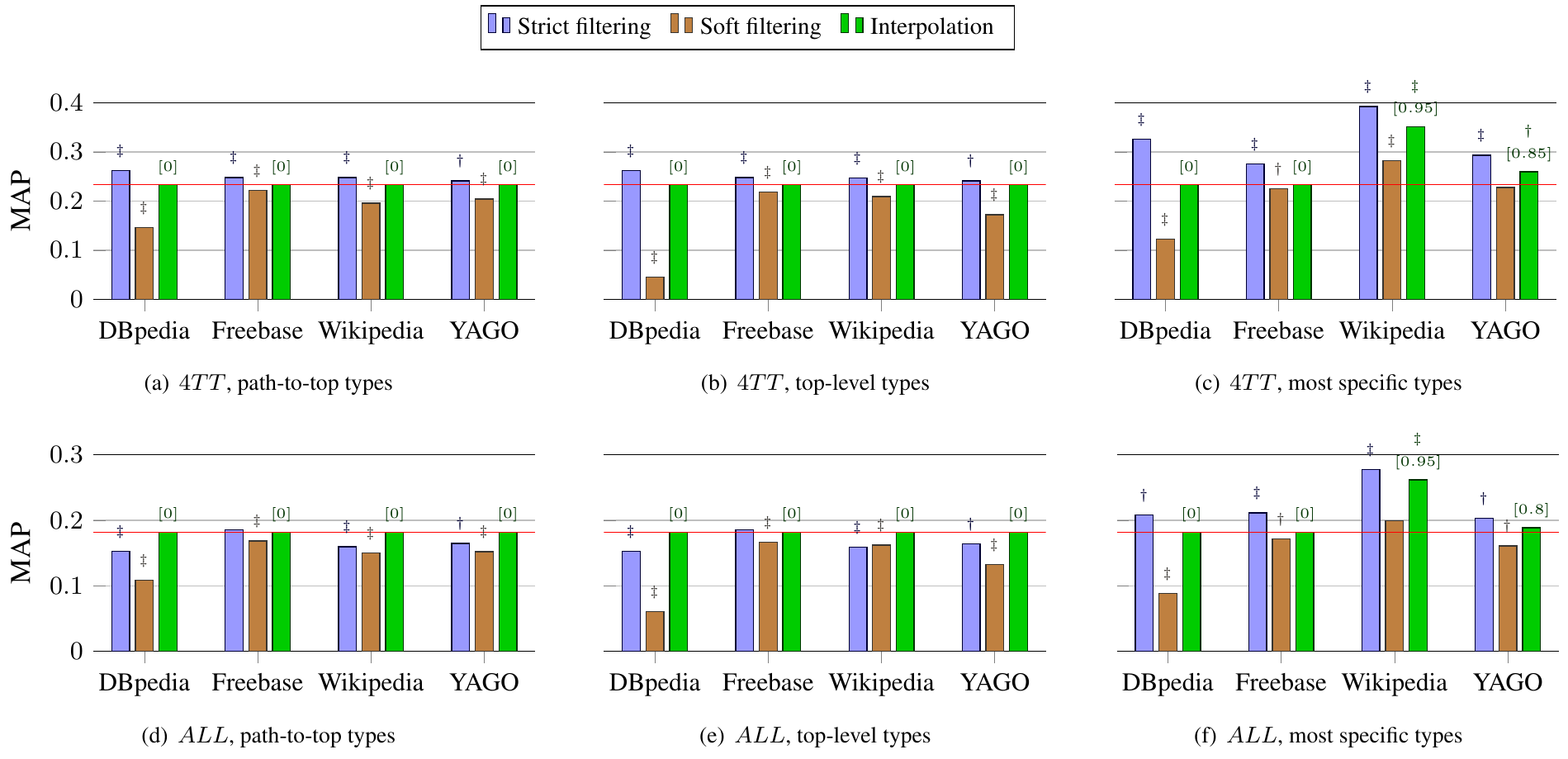}
    \centering
    \caption{Entity retrieval performance for all combinations of type taxonomies, type representation modes, and retrieval models. (Top): only entities with types from all four type taxonomies; (Bottom): all entities in the knowledge base. The red line corresponds to the term-based baseline.    Above each bar, the symbols $^\dag$ and $^\ddag$ indicate statistical significance against the baseline; the numbers in brackets show the type weight (empirically found best $\lambda_t$) used by the interpolation model.}
    \label{fig:res:end2end_results}
\end{figure*}

In this section we present evaluation results for all combinations of the three proposed dimensions: type taxonomies, type representation modes, and retrieval models.  When discussing the results, we use the term \emph{configuration} to refer to a particular combination of type taxonomy, type representation, and retrieval model.

Figure~\ref{fig:res:end2end_results} shows the results, corresponding to the two settings we distinguished in Section~\ref{sec:expsetup:allVS4tt}:
in the top histograms, we consider only entities that have types assigned to them in all four type taxonomies (4TT); in the bottom histograms, we rank all entities in the knowledge base (ALL).  The red line corresponds to the term-based baseline.
Our evaluation metric is Mean Average Precision (MAP).  
We report on statistical significance using a two-tailed paired t-test at $p<0.05$ and $p<0.001$, denoted by $^\dag$ and $^\ddag$, respectively.

\paragraph{RQ1.}
Let us turn to our first research question, which concerns the impact of the particular choice of type taxonomy.  It is clear that Wikipedia, in combination with the most specific type representation, performs best (for both 4TT and ALL). In particular, for the 4TT setting (top right plot in Figure~\ref{fig:res:end2end_results}), the improvements for Wikipedia are highly significant for all three retrieval models.  
As for the rest, there is no easy way to compare taxonomies, as the performance varies depending on the other dimensions. E.g., for 4TT using strict filtering and more general types (i.e., the purple bars in the top left and top middle histograms in Figure~\ref{fig:res:end2end_results}), the smaller, shallower type taxonomies (DBpedia and Freebase) tend to outperform the larger, deeper ones (Wikipedia and YAGO).

\paragraph{RQ2.}
The second research question, which is about type representation, has a clear answer: keeping only the most specific types in the hierarchy provides the best performance (right vs. left and middle histograms in Figure~\ref{fig:res:end2end_results}).  This is also in line with findings in past work (cf. Section~\ref{sec:rel}).  
As for the other two representations, i.e., types along path to top vs. top-level types, two things are worth pointing out. 
Firstly, the results are the same for both type representations when using strict filtering, which is explained by how the representations are defined in Section~\ref{sec:dim_reprs}; if an entity is retained (given that the intersection between the entity's types and the target types is non-empty), this filtering does not change by adding more specific types.  
Secondly, for the interpolation model, we can observe that the $\lambda_t$ weights are always $0$ for these representations.  This means that type information is not used at all.  
Overall, we could not find any evidence that hierarchical relationships from ancestor types would benefit retrieval effectiveness.  

\paragraph{RQ3.}  
Answering our final research question, concerning the type-aware retrieval model, requires a more elaborate treatment.
In the 4TT setting, strict filtering is the best retrieval model for every configuration, outperforming the baseline with high significance in almost all cases.  This no longer holds in the ALL setting; in fact, all MAP scores drop with respect to the corresponding 4TT configuration.
This is expected, as in the more realistic setting, many relevant entities may have incomplete type assignments.  Only the interpolation model can deal with this in a robust manner.

\begin{figure*}[!th]
    \centering
    \includegraphics[width=0.7\textwidth]{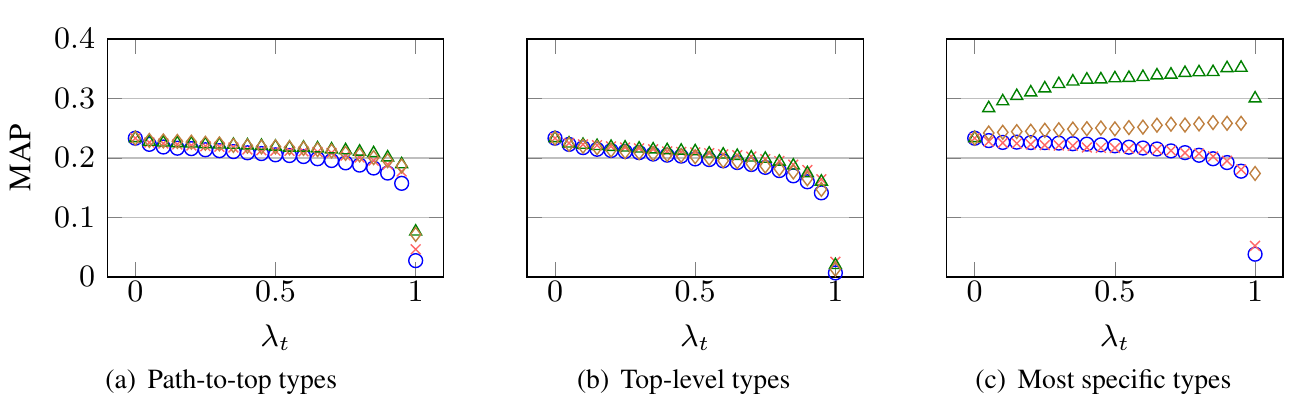}
    \hspace{0.3in}\includegraphics[width=0.09\textwidth]{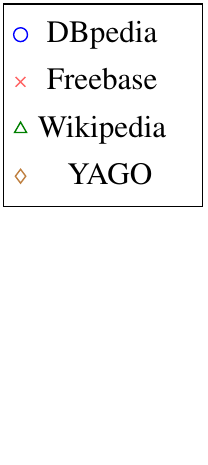}  
    \centering
    \vspace*{-\baselineskip}
    \caption{Retrieval performance (4TT setting) using the interpolation model with different type weights, $\lambda_{t}$.}
    \label{fig:res:inter_lambda_vs_map}
\end{figure*}

Figure~\ref{fig:res:inter_lambda_vs_map} shows the performance of the interpolation model when varying the weight of the type-based component (value of $\lambda_t$).  Due to space constraints, we present the plots only for the 4TT setting; the figures look very similar for the ALL setting.
We find that for the smaller, shallower type taxonomies, DBpedia and Freebase, assigning more weight to type-based information is increasingly more harmful, independently of the type representation or type assignment setting.  The same occurs for Wikipedia and YAGO using the more general type representations.  On the other hand, when using only the most specific types (right plot in Figure~\ref{fig:res:inter_lambda_vs_map}), for Wikipedia and YAGO, performance increases with higher $\lambda_t$ values.  Yet, MAP scores peak at $\lambda_t<1$, meaning that term-based similarity is still needed for optimal performance.

The only configurations performing worse than the baseline, even in the 4TT setting, are the ones using the soft filtering model.  In particular, the MAP scores for DBpedia with soft filtering are noticeably low.  We plan to perform a deeper investigation of this phenomenon in future work.

%% file: ictir2017-types-08.tex
\section{Analysis and Discussion}
\label{sec:analysis}

Now that we have presented our results, we proceed with further analysis of some of the issues we identified in the previous section.

\subsection{Missing Type Information}
\label{sec:analysis:missing}

\begin{table}[!t]
  \centering
  \caption{Number of entities with missing type information.}
  \label{table:discussion:oracle}
  \begin{tabular}{ l r r r r}
    \toprule
    Entities & DBpedia & Freebase & Wikipedia & YAGO \\
    \midrule
    All entities & 35,390 & 1,369,636 & 1,113,299 & 1,755,480 \\
    Relevant entities & 3,341 & 1,594 & 2,567 & 2,532 \\
    \bottomrule
  \end{tabular}
\end{table}
 
In order to make a fair comparison between different type taxonomies, we had to account for the fact that the entity type assignments in the knowledge bases may be incomplete (cf. the 4TT setting in Section~\ref{sec:expsetup:allVS4tt}).  Indeed, results in Section~\ref{sec:results} have shown that the benefits of using type information are more obvious when entities are not missing type assignments.
Table~\ref{table:discussion:oracle} shows, for each of the type taxonomies, the number of entities that have no types assigned to them in the KB (i.e., ``non-typed'' entities).  Interestingly, while DBpedia has the least number of non-typed entities (only 35K out of 4.6M), it lacks types for over 25\% of the relevant entities (3.3K out of 12.9K).
Even for Freebase, which has the best coverage of relevant entities, over 12\% of the relevant entities have no type assignments in the KB.
Clearly, the problem of missing type information, frequently referred to as partial \emph{extensional} coverage of type systems~\citep{Gangemi:2012:ATD}, is an important area of research (cf. Section~\ref{sec:rel}).

\subsection{Revisiting the Target Types Oracle}
\label{sec:analysis:oracle}

\begin{figure*}[!th]
    \centering
    \includegraphics[width=1.0\textwidth]{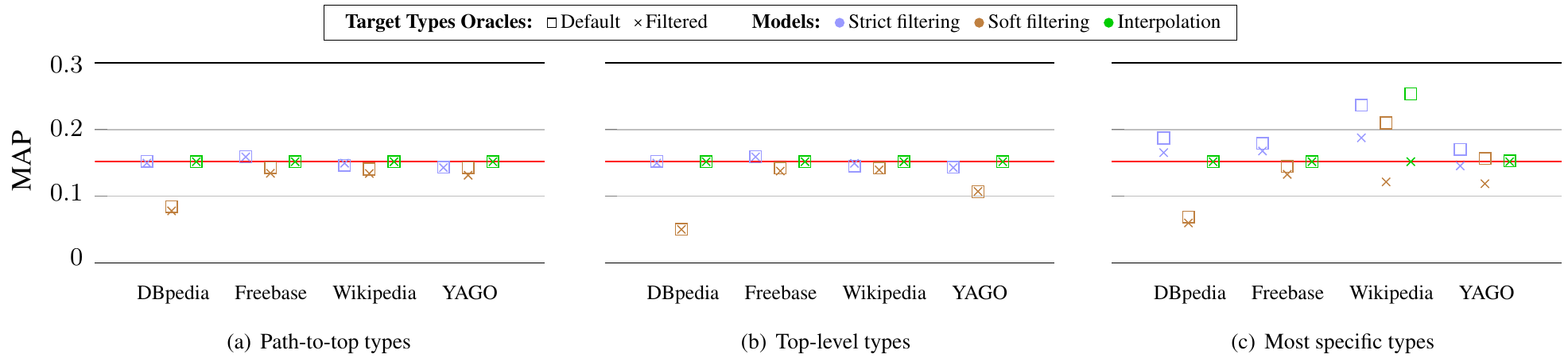}
    \centering
    \caption{Retrieval performance using the default vs. filtered target types oracle.  The red line is the term-based baseline.}
    \label{fig:discussion:oracle}
\end{figure*}

Another aspect of type-based information we are concerned about is the quality of target types.  Previously, we have included all types associated with known relevant entities, proportional to their frequency, in the target type distribution ($\theta_q$); we shall refer to it as the \emph{default oracle}.
Here, we consider another variant, referred to as \emph{filtered oracle}, where a frequency threshold is applied.  Specifically, we include type $t$ as a target type iff at least 3 relevant entities have $t$ assigned to them.
As a consequence of this filtering, many queries have an empty set of types; for this experiment, we discard those from the ground truth set, leaving us with 182 queries in total. 

A comparison of the two oracles is presented in Figure~\ref{fig:discussion:oracle}.  
For the more general type representations, the filtered oracle turns out to be slightly less effective for most of the configurations.  Yet, the differences are barely noticeable. 
When using the most specific types, we find that MAP scores drop, especially for larger, deeper taxonomies (Wikipedia and YAGO); some configurations no longer outperform the term-based baseline.
Hence, it is important to consider all possible target types, even those with a low probability.

\subsection{What is in a Target Type?}
\label{sec:analysis:hgtt}

Our ultimate interest in this work is in understanding the usefulness of type information for ad-hoc entity retrieval.  What portion of relevant entities can target types help to capture?  To shed some light on this, we measure the coverage of relevant entities by (i) the top ranked type and (ii) the set of top 3 types.\footnote{For this experiment, we take the type assignments ``as-recorded'' in Wikipedia, without enforcing the taxonomical constraints.}
Table~\ref{table:discussion:wiki_tags_full} reports the results.  We find that Wikipedia has, by far, the highest precision and F1-score among all type taxonomies; YAGO comes second.  Notice that these are the two taxonomies that performed best, when using the most specific type representations, in Figure~\ref{fig:res:end2end_results}. 

In summary, we have found that specific types from large, fine-grained taxonomies provide the best performance.  Yet, it appears that it is not the hierarchical nature of the taxonomy that brings benefits, but rather the fact that these fine-grained types provide semantic sets or classes that can capture (some subset of) the relevant entities with high precision.

\begin{table}[!t]
  \centering
  \caption{Coverage of relevant entities by top-$K$ types, in terms of precision, recall, and F1, averaged over all queries.}
  \label{table:discussion:wiki_tags_full}
  \begin{tabular}{ l l r r r}
    \toprule
    Top-$K$ types & Type Taxonomy & \multicolumn{1}{c}{P} & \multicolumn{1}{c}{R} & \multicolumn{1}{c}{F1} \\  
    \midrule
    $K$ = 1 & DBpedia & 0.0027 & 0.5863 & 0.0046 \\
    & Freebase & 0.0060 & \textbf{0.7254} & 0.0076 \\
    & Wikipedia & \textbf{0.1147} & 0.4798 & \textbf{0.1287} \\
    & YAGO & 0.0418 & 0.6303 & 0.0488 \\
    \midrule
    $K$ = 3 & DBpedia & 0.0006 & 0.7199 & 0.0012 \\
    & Freebase & 0.0004 & \textbf{0.7805} & 0.0008 \\
    & Wikipedia & \textbf{0.0402} & 0.5847 & \textbf{0.0614} \\
    & YAGO & 0.0036 & 0.7025 & 0.0062 \\
    \bottomrule
  \end{tabular}
  \vspace{-0.05in}  
\end{table}

%% file: ictir2017-types-09.tex
\section{Conclusions}
\label{sec:concl}

In this paper we have furthered our understanding on the usage of target type information for entity retrieval over structured data sources.  
A main contribution of this work is the systematic comparison of four well-known type taxonomies (DBpedia, Freebase, Wikipedia, and YAGO) across three dimensions of interest: the representation of hierarchical entity type information, the way to combine term-based and type-based information, and the impact of choosing a particular type taxonomy.  
We have found that using the most specific types in a fine-grained taxonomy, like Wikipedia, leads to the best retrieval effectiveness.

We identify two directions for future work.  First, we plan to report on an even deeper query-level analysis, which was not possible here due to space limitations.
Second, our investigations so far have taken place in an idealized environment, assuming that an ``oracle'' process can provide us with the target types for each query.  We wish to perform a similar analysis using automatically identified target types~\cite{Garigliotti:2017:TTI}.

%% file: 00paper.bbl

\begin{thebibliography}{00}


\ifx \showCODEN    \undefined \def \showCODEN     #1{\unskip}     \fi
\ifx \showDOI      \undefined \def \showDOI       #1{{\tt DOI:}\penalty0{#1}\ }
  \fi
\ifx \showISBNx    \undefined \def \showISBNx     #1{\unskip}     \fi
\ifx \showISBNxiii \undefined \def \showISBNxiii  #1{\unskip}     \fi
\ifx \showISSN     \undefined \def \showISSN      #1{\unskip}     \fi
\ifx \showLCCN     \undefined \def \showLCCN      #1{\unskip}     \fi
\ifx \shownote     \undefined \def \shownote      #1{#1}          \fi
\ifx \showarticletitle \undefined \def \showarticletitle #1{#1}   \fi
\ifx \showURL      \undefined \def \showURL       #1{#1}          \fi
\providecommand\bibfield[2]{#2}
\providecommand\bibinfo[2]{#2}
\providecommand\natexlab[1]{#1}
\providecommand\showeprint[2][]{arXiv:#2}

\bibitem[\protect\citeauthoryear{Balog, Bron, and {De Rijke}}{Balog
  et~al\mbox{.}}{2011}]%
        {Balog:2011:QME}
\bibfield{author}{\bibinfo{person}{Krisztian Balog}, \bibinfo{person}{Marc
  Bron}, {and} \bibinfo{person}{Maarten {De Rijke}}.}
  \bibinfo{year}{2011}\natexlab{}.
\newblock \showarticletitle{Query modeling for entity search based on terms,
  categories, and examples}.
\newblock \bibinfo{journal}{{\em ACM Trans. Inf. Syst.\/}}
  \bibinfo{volume}{29}, \bibinfo{number}{4} (\bibinfo{year}{2011}),
  \bibinfo{pages}{22:1--22:31}.
\newblock


\bibitem[\protect\citeauthoryear{Balog, de~Vries, Serdyukov, Thomas, and
  Westerveld}{Balog et~al\mbox{.}}{2010}]%
        {Balog:2010:OTE}
\bibfield{author}{\bibinfo{person}{K. Balog}, \bibinfo{person}{A.~P. de Vries},
  \bibinfo{person}{P. Serdyukov}, \bibinfo{person}{P. Thomas}, {and}
  \bibinfo{person}{T. Westerveld}.} \bibinfo{year}{2010}\natexlab{}.
\newblock \showarticletitle{Overview of the {TREC} 2009 Entity Track}. In
  \bibinfo{booktitle}{{\em Proc. of TREC}}.
\newblock


\bibitem[\protect\citeauthoryear{Balog and Neumayer}{Balog and
  Neumayer}{2012}]%
        {Balog:2012:HTT}
\bibfield{author}{\bibinfo{person}{Krisztian Balog} {and}
  \bibinfo{person}{Robert Neumayer}.} \bibinfo{year}{2012}\natexlab{}.
\newblock \showarticletitle{Hierarchical target type identification for
  entity-oriented queries}. In \bibinfo{booktitle}{{\em Proc. of CIKM}}.
  \bibinfo{pages}{2391--2394}.
\newblock


\bibitem[\protect\citeauthoryear{Balog and Neumayer}{Balog and
  Neumayer}{2013}]%
        {Balog:2013:TCE}
\bibfield{author}{\bibinfo{person}{Krisztian Balog} {and}
  \bibinfo{person}{Robert Neumayer}.} \bibinfo{year}{2013}\natexlab{}.
\newblock \showarticletitle{{A Test Collection for Entity Search in DBpedia}}.
  In \bibinfo{booktitle}{{\em Proc. of SIGIR}}. \bibinfo{pages}{737--740}.
\newblock


\bibitem[\protect\citeauthoryear{Balog, Serdyukov, and De~Vries}{Balog
  et~al\mbox{.}}{2012}]%
        {Balog:2012:OTE}
\bibfield{author}{\bibinfo{person}{Krisztian Balog}, \bibinfo{person}{Pavel
  Serdyukov}, {and} \bibinfo{person}{Arjen~P. De~Vries}.}
  \bibinfo{year}{2012}\natexlab{}.
\newblock \showarticletitle{Overview of the {TREC} 2011 Entity Track}. In
  \bibinfo{booktitle}{{\em Proc. of TREC}}.
\newblock


\bibitem[\protect\citeauthoryear{Bron, Balog, and de~Rijke}{Bron
  et~al\mbox{.}}{2010}]%
        {Bron:2010:RRE}
\bibfield{author}{\bibinfo{person}{Marc Bron}, \bibinfo{person}{Krisztian
  Balog}, {and} \bibinfo{person}{Maarten de Rijke}.}
  \bibinfo{year}{2010}\natexlab{}.
\newblock \showarticletitle{Ranking Related Entities: Components and Analyses}.
  In \bibinfo{booktitle}{{\em Proc. of CIKM}}. \bibinfo{pages}{1079--1088}.
\newblock


\bibitem[\protect\citeauthoryear{Demartini, Firan, and Iofciu}{Demartini
  et~al\mbox{.}}{2008}]%
        {Demartini:2008:LIQ}
\bibfield{author}{\bibinfo{person}{Gianluca Demartini},
  \bibinfo{person}{Claudiu~S. Firan}, {and} \bibinfo{person}{Tereza Iofciu}.}
  \bibinfo{year}{2008}\natexlab{}.
\newblock \showarticletitle{Focused Access to XML Documents}.
\newblock \bibinfo{publisher}{Springer}, Chapter L3S at INEX 2007,
  \bibinfo{pages}{252--263}.
\newblock
\showISBNx{978-3-540-85901-7}


\bibitem[\protect\citeauthoryear{Demartini, Firan, Iofciu, Krestel, and
  Nejdl}{Demartini et~al\mbox{.}}{2010a}]%
        {Demartini:2010:WFE}
\bibfield{author}{\bibinfo{person}{Gianluca Demartini},
  \bibinfo{person}{Claudiu~S. Firan}, \bibinfo{person}{Tereza Iofciu},
  \bibinfo{person}{Ralf Krestel}, {and} \bibinfo{person}{Wolfgang Nejdl}.}
  \bibinfo{year}{2010}\natexlab{a}.
\newblock \showarticletitle{Why finding entities in Wikipedia is difficult,
  sometimes}.
\newblock \bibinfo{journal}{{\em Information Retrieval\/}}
  \bibinfo{volume}{13}, \bibinfo{number}{5} (\bibinfo{date}{may}
  \bibinfo{year}{2010}), \bibinfo{pages}{534--567}.
\newblock
\showISSN{1386-4564}


\bibitem[\protect\citeauthoryear{Demartini, Iofciu, and De~Vries}{Demartini
  et~al\mbox{.}}{2010b}]%
        {Demartini:2009:OIE}
\bibfield{author}{\bibinfo{person}{Gianluca Demartini}, \bibinfo{person}{Tereza
  Iofciu}, {and} \bibinfo{person}{Arjen~P. De~Vries}.}
  \bibinfo{year}{2010}\natexlab{b}.
\newblock \showarticletitle{Overview of the INEX 2009 Entity Ranking Track}.
\newblock In \bibinfo{booktitle}{{\em Focused Retrieval and Evaluation, and
  INEX}}. \bibinfo{pages}{254--264}.
\newblock


\bibitem[\protect\citeauthoryear{Demartini, Iofciu, and De~Vries}{Demartini
  et~al\mbox{.}}{2010c}]%
        {Demartini:2010:OIE}
\bibfield{author}{\bibinfo{person}{Gianluca Demartini}, \bibinfo{person}{Tereza
  Iofciu}, {and} \bibinfo{person}{Arjen~P. De~Vries}.}
  \bibinfo{year}{2010}\natexlab{c}.
\newblock \showarticletitle{{Overview of the INEX 2009 Entity Ranking Track}}.
\newblock In \bibinfo{booktitle}{{\em Focused Retrieval and Evaluation}}.
  \bibinfo{pages}{254--264}.
\newblock


\bibitem[\protect\citeauthoryear{Fleischman and Hovy}{Fleischman and
  Hovy}{2002}]%
        {Fleischman:2002:FGC}
\bibfield{author}{\bibinfo{person}{Michael Fleischman} {and}
  \bibinfo{person}{Eduard Hovy}.} \bibinfo{year}{2002}\natexlab{}.
\newblock \showarticletitle{Fine Grained Classification of Named Entities}. In
  \bibinfo{booktitle}{{\em Proc. of COLING}}. \bibinfo{pages}{1--7}.
\newblock


\bibitem[\protect\citeauthoryear{Fossati, Kontokostas, and Lehmann}{Fossati
  et~al\mbox{.}}{2015}]%
        {Fossati:2015:ULE}
\bibfield{author}{\bibinfo{person}{Marco Fossati}, \bibinfo{person}{Dimitris
  Kontokostas}, {and} \bibinfo{person}{Jens Lehmann}.}
  \bibinfo{year}{2015}\natexlab{}.
\newblock \showarticletitle{Unsupervised Learning of an Extensive and Usable
  Taxonomy for DBpedia}. In \bibinfo{booktitle}{{\em Proc. of SEMANTICS}}.
  \bibinfo{pages}{177--184}.
\newblock


\bibitem[\protect\citeauthoryear{Gangemi, Nuzzolese, Presutti, Draicchio,
  Musetti, and Ciancarini}{Gangemi et~al\mbox{.}}{2012}]%
        {Gangemi:2012:ATD}
\bibfield{author}{\bibinfo{person}{Aldo Gangemi},
  \bibinfo{person}{Andrea~Giovanni Nuzzolese}, \bibinfo{person}{Valentina
  Presutti}, \bibinfo{person}{Francesco Draicchio}, \bibinfo{person}{Alberto
  Musetti}, {and} \bibinfo{person}{Paolo Ciancarini}.}
  \bibinfo{year}{2012}\natexlab{}.
\newblock \showarticletitle{Automatic Typing of DBpedia Entities}. In
  \bibinfo{booktitle}{{\em Proc. of ISWC}}. \bibinfo{pages}{65--81}.
\newblock


\bibitem[\protect\citeauthoryear{Garigliotti, Hasibi, and Balog}{Garigliotti
  et~al\mbox{.}}{2017}]%
        {Garigliotti:2017:TTI}
\bibfield{author}{\bibinfo{person}{Dar\'{\i}o Garigliotti},
  \bibinfo{person}{Faegheh Hasibi}, {and} \bibinfo{person}{Krisztian Balog}.}
  \bibinfo{year}{2017}\natexlab{}.
\newblock \showarticletitle{Target Type Identification for Entity-Bearing
  Queries}. In \bibinfo{booktitle}{{\em Proc. of SIGIR}}.
  \bibinfo{pages}{845--848}.
\newblock


\bibitem[\protect\citeauthoryear{Giuliano}{Giuliano}{2009}]%
        {Giuliano:2009:FCN}
\bibfield{author}{\bibinfo{person}{Claudio Giuliano}.}
  \bibinfo{year}{2009}\natexlab{}.
\newblock \showarticletitle{Fine-grained Classification of Named Entities
  Exploiting Latent Semantic Kernels}. In \bibinfo{booktitle}{{\em Proc. of
  CoNLL}}. \bibinfo{pages}{201--209}.
\newblock


\bibitem[\protect\citeauthoryear{J\"{a}msen, N\"{a}ppil\"{a}, and
  Arvola}{J\"{a}msen et~al\mbox{.}}{2008}]%
        {Jamsen:2008:ERB}
\bibfield{author}{\bibinfo{person}{Janne J\"{a}msen}, \bibinfo{person}{Turkka
  N\"{a}ppil\"{a}}, {and} \bibinfo{person}{Paavo Arvola}.}
  \bibinfo{year}{2008}\natexlab{}.
\newblock \showarticletitle{Focused Access to XML Documents}.
\newblock \bibinfo{publisher}{Springer}, Chapter Entity Ranking Based on
  Category Expansion, \bibinfo{pages}{264--278}.
\newblock
\showISBNx{978-3-540-85901-7}


\bibitem[\protect\citeauthoryear{Kaptein and Kamps}{Kaptein and Kamps}{2009}]%
        {Kaptein:2009:FEW}
\bibfield{author}{\bibinfo{person}{Rianne Kaptein} {and} \bibinfo{person}{Jaap
  Kamps}.} \bibinfo{year}{2009}\natexlab{}.
\newblock \showarticletitle{Finding Entities in Wikipedia using Links and
  Categories}. In \bibinfo{booktitle}{{\em Advances in Focused Retrieval,
  INEX}}. \bibinfo{pages}{273--279}.
\newblock


\bibitem[\protect\citeauthoryear{Kaptein and Kamps}{Kaptein and Kamps}{2013}]%
        {Kaptein:2013:ECS}
\bibfield{author}{\bibinfo{person}{Rianne Kaptein} {and} \bibinfo{person}{Jaap
  Kamps}.} \bibinfo{year}{2013}\natexlab{}.
\newblock \showarticletitle{Exploiting the category structure of Wikipedia for
  entity ranking}.
\newblock \bibinfo{journal}{{\em Artificial Intelligence\/}}
  \bibinfo{volume}{194} (\bibinfo{date}{jan} \bibinfo{year}{2013}),
  \bibinfo{pages}{111--129}.
\newblock
\showISSN{00043702}


\bibitem[\protect\citeauthoryear{Kaptein, Serdyukov, {De Vries}, and
  Kamps}{Kaptein et~al\mbox{.}}{2010}]%
        {Kaptein:2010:ERU}
\bibfield{author}{\bibinfo{person}{Rianne Kaptein}, \bibinfo{person}{Pavel
  Serdyukov}, \bibinfo{person}{Arjen~P. {De Vries}}, {and}
  \bibinfo{person}{Jaap Kamps}.} \bibinfo{year}{2010}\natexlab{}.
\newblock \showarticletitle{Entity ranking using Wikipedia as a pivot}. In
  \bibinfo{booktitle}{{\em Proc. of CIKM}}. \bibinfo{pages}{69--78}.
\newblock


\bibitem[\protect\citeauthoryear{Lehmann, Isele, Jakob, Jentzsch, Kontokostas,
  Mendes, Hellmann, Morsey, van Kleef, Auer, and Bizer}{Lehmann
  et~al\mbox{.}}{2015}]%
        {Lehmann:2015:DAL}
\bibfield{author}{\bibinfo{person}{Jens Lehmann}, \bibinfo{person}{Robert
  Isele}, \bibinfo{person}{Max Jakob}, \bibinfo{person}{Anja Jentzsch},
  \bibinfo{person}{Dimitris Kontokostas}, \bibinfo{person}{Pablo~N. Mendes},
  \bibinfo{person}{Sebastian Hellmann}, \bibinfo{person}{Mohamed Morsey},
  \bibinfo{person}{Patrick van Kleef}, \bibinfo{person}{S{\"{o}}ren Auer},
  {and} \bibinfo{person}{Christian Bizer}.} \bibinfo{year}{2015}\natexlab{}.
\newblock \showarticletitle{DBpedia - {A} large-scale, multilingual knowledge
  base extracted from Wikipedia}.
\newblock \bibinfo{journal}{{\em Semantic Web\/}} \bibinfo{volume}{6},
  \bibinfo{number}{2} (\bibinfo{year}{2015}), \bibinfo{pages}{167--195}.
\newblock


\bibitem[\protect\citeauthoryear{Lin, Mausam, and Etzioni}{Lin
  et~al\mbox{.}}{2012}]%
        {Lin:2012:NNP}
\bibfield{author}{\bibinfo{person}{Thomas Lin}, \bibinfo{person}{Mausam}, {and}
  \bibinfo{person}{Oren Etzioni}.} \bibinfo{year}{2012}\natexlab{}.
\newblock \showarticletitle{No Noun Phrase Left Behind: Detecting and Typing
  Unlinkable Entities}. In \bibinfo{booktitle}{{\em Proc. of EMNLP-CoNLL}}.
  \bibinfo{pages}{893--903}.
\newblock


\bibitem[\protect\citeauthoryear{Ling and Weld}{Ling and Weld}{2012}]%
        {Ling:2012:FER}
\bibfield{author}{\bibinfo{person}{Xiao Ling} {and} \bibinfo{person}{Daniel~S.
  Weld}.} \bibinfo{year}{2012}\natexlab{}.
\newblock \showarticletitle{Fine-grained Entity Recognition}. In
  \bibinfo{booktitle}{{\em Proc. of AAAI}}. \bibinfo{pages}{94--100}.
\newblock


\bibitem[\protect\citeauthoryear{Lopez, Unger, Cimiano, and Motta}{Lopez
  et~al\mbox{.}}{2013}]%
        {Lopez:2013:EQA}
\bibfield{author}{\bibinfo{person}{Vanessa Lopez}, \bibinfo{person}{Christina
  Unger}, \bibinfo{person}{Philipp Cimiano}, {and} \bibinfo{person}{Enrico
  Motta}.} \bibinfo{year}{2013}\natexlab{}.
\newblock \showarticletitle{Evaluating Question Answering over Linked Data}.
\newblock \bibinfo{journal}{{\em Web Semantics: Science, Services and Agents on
  the World Wide Web\/}}  \bibinfo{volume}{21} (\bibinfo{date}{aug}
  \bibinfo{year}{2013}), \bibinfo{pages}{3--13}.
\newblock
\showISSN{1570-8268}


\bibitem[\protect\citeauthoryear{Mika}{Mika}{2013}]%
        {Mika:2013:ESW}
\bibfield{author}{\bibinfo{person}{Peter Mika}.}
  \bibinfo{year}{2013}\natexlab{}.
\newblock \showarticletitle{{Entity Search on the Web}}. In
  \bibinfo{booktitle}{{\em Proc. of WWW}}. \bibinfo{pages}{1231--1232}.
\newblock


\bibitem[\protect\citeauthoryear{Nakashole, Tylenda, and Weikum}{Nakashole
  et~al\mbox{.}}{2013}]%
        {Nakashole:2013:FST}
\bibfield{author}{\bibinfo{person}{Ndapandula Nakashole},
  \bibinfo{person}{Tomasz Tylenda}, {and} \bibinfo{person}{Gerhard Weikum}.}
  \bibinfo{year}{2013}\natexlab{}.
\newblock \showarticletitle{Fine-grained Semantic Typing of Emerging Entities}.
  In \bibinfo{booktitle}{{\em Proc. of ACL}}. \bibinfo{pages}{1488--1497}.
\newblock


\bibitem[\protect\citeauthoryear{Neumayer, Balog, and N{\o}rv{\aa}g}{Neumayer
  et~al\mbox{.}}{2012a}]%
        {Neumayer:2012:MEA}
\bibfield{author}{\bibinfo{person}{Robert Neumayer}, \bibinfo{person}{Krisztian
  Balog}, {and} \bibinfo{person}{Kjetil N{\o}rv{\aa}g}.}
  \bibinfo{year}{2012}\natexlab{a}.
\newblock \showarticletitle{On the modeling of entities for ad-hoc entity
  search in the web of data}. In \bibinfo{booktitle}{{\em Proc. of ECIR}}.
  \bibinfo{pages}{133--145}.
\newblock


\bibitem[\protect\citeauthoryear{Neumayer, Balog, and N{\o}rv{\aa}g}{Neumayer
  et~al\mbox{.}}{2012b}]%
        {Neumayer:2012:SGE}
\bibfield{author}{\bibinfo{person}{Robert Neumayer}, \bibinfo{person}{Krisztian
  Balog}, {and} \bibinfo{person}{Kjetil N{\o}rv{\aa}g}.}
  \bibinfo{year}{2012}\natexlab{b}.
\newblock \showarticletitle{When simple is (more than) good enough: effective
  semantic search with (almost) no semantics}. In \bibinfo{booktitle}{{\em
  Proc. of ECIR}}. \bibinfo{pages}{540--543}.
\newblock


\bibitem[\protect\citeauthoryear{Nuzzolese, Gangemi, Presutti, and
  Ciancarini}{Nuzzolese et~al\mbox{.}}{2012}]%
        {Nuzzolese:2012:TIT}
\bibfield{author}{\bibinfo{person}{Andrea~Giovanni Nuzzolese},
  \bibinfo{person}{Aldo Gangemi}, \bibinfo{person}{Valentina Presutti}, {and}
  \bibinfo{person}{Paolo Ciancarini}.} \bibinfo{year}{2012}\natexlab{}.
\newblock \showarticletitle{{Type inference through the analysis of Wikipedia
  links}}. In \bibinfo{booktitle}{{\em Proc. of LDOW}}.
\newblock


\bibitem[\protect\citeauthoryear{Pehcevski, Thom, Vercoustre, and
  Naumovski}{Pehcevski et~al\mbox{.}}{2010}]%
        {Pehcevski:2010:ERW}
\bibfield{author}{\bibinfo{person}{Jovan Pehcevski}, \bibinfo{person}{James~A
  Thom}, \bibinfo{person}{Anne-Marie Vercoustre}, {and}
  \bibinfo{person}{Vladimir Naumovski}.} \bibinfo{year}{2010}\natexlab{}.
\newblock \showarticletitle{Entity ranking in Wikipedia: utilising categories,
  links and topic difficulty prediction}.
\newblock \bibinfo{journal}{{\em Information Retrieval\/}}
  \bibinfo{volume}{13}, \bibinfo{number}{5} (\bibinfo{year}{2010}),
  \bibinfo{pages}{568--600}.
\newblock
\showISSN{13864564}


\bibitem[\protect\citeauthoryear{Pound, Mika, and Zaragoza}{Pound
  et~al\mbox{.}}{2010}]%
        {Pound:2010:AOR}
\bibfield{author}{\bibinfo{person}{Jeffrey Pound}, \bibinfo{person}{Peter
  Mika}, {and} \bibinfo{person}{Hugo Zaragoza}.}
  \bibinfo{year}{2010}\natexlab{}.
\newblock \showarticletitle{Ad-hoc object retrieval in the web of data}. In
  \bibinfo{booktitle}{{\em Proc. of WWW}}. \bibinfo{pages}{771--780}.
\newblock


\bibitem[\protect\citeauthoryear{Rahman and Ng}{Rahman and Ng}{2010}]%
        {Rahman:2010:IFS}
\bibfield{author}{\bibinfo{person}{Altaf Rahman} {and} \bibinfo{person}{Vincent
  Ng}.} \bibinfo{year}{2010}\natexlab{}.
\newblock \showarticletitle{Inducing Fine-grained Semantic Classes via
  Hierarchical and Collective Classification}. In \bibinfo{booktitle}{{\em
  Proc. of COLING}}. \bibinfo{pages}{931--939}.
\newblock


\bibitem[\protect\citeauthoryear{Raviv, Carmel, and Kurland}{Raviv
  et~al\mbox{.}}{2012}]%
        {Raviv:2012:RFE}
\bibfield{author}{\bibinfo{person}{Hadas Raviv}, \bibinfo{person}{David
  Carmel}, {and} \bibinfo{person}{Oren Kurland}.}
  \bibinfo{year}{2012}\natexlab{}.
\newblock \showarticletitle{A Ranking Framework for Entity Oriented Search
  Using Markov Random Fields}. In \bibinfo{booktitle}{{\em Proc. of JIWES}}.
  \bibinfo{pages}{1:1--1:6}.
\newblock


\bibitem[\protect\citeauthoryear{Sawant and Chakrabarti}{Sawant and
  Chakrabarti}{2013}]%
        {Sawant:2013:LJQ}
\bibfield{author}{\bibinfo{person}{Uma Sawant} {and} \bibinfo{person}{S
  Chakrabarti}.} \bibinfo{year}{2013}\natexlab{}.
\newblock \showarticletitle{Learning Joint Query Interpretation and Response
  Ranking}. In \bibinfo{booktitle}{{\em Proc. of WWW}}.
  \bibinfo{pages}{1099--1109}.
\newblock


\bibitem[\protect\citeauthoryear{Suchanek, Kasneci, and Weikum}{Suchanek
  et~al\mbox{.}}{2007}]%
        {Suchanek:2007:YCS}
\bibfield{author}{\bibinfo{person}{Fabian~M Suchanek}, \bibinfo{person}{Gjergji
  Kasneci}, {and} \bibinfo{person}{Gerhard Weikum}.}
  \bibinfo{year}{2007}\natexlab{}.
\newblock \showarticletitle{Yago: A Core of Semantic Knowledge}. In
  \bibinfo{booktitle}{{\em Proc. of WWW}}. \bibinfo{pages}{697--706}.
\newblock


\bibitem[\protect\citeauthoryear{Tonon, Catasta, Demartini,
  Cudr{\'{e}}-Mauroux, and Aberer}{Tonon et~al\mbox{.}}{2013}]%
        {Tonon:2013:TRE}
\bibfield{author}{\bibinfo{person}{Alberto Tonon}, \bibinfo{person}{Michele
  Catasta}, \bibinfo{person}{Gianluca Demartini}, \bibinfo{person}{Philippe
  Cudr{\'{e}}-Mauroux}, {and} \bibinfo{person}{Karl Aberer}.}
  \bibinfo{year}{2013}\natexlab{}.
\newblock \showarticletitle{TRank: Ranking Entity Types Using the Web of Data}.
  In \bibinfo{booktitle}{{\em Proc. of ISWC}}. \bibinfo{pages}{640--656}.
\newblock


\bibitem[\protect\citeauthoryear{Vallet and Zaragoza}{Vallet and
  Zaragoza}{2008}]%
        {Vallet:2008:IMI}
\bibfield{author}{\bibinfo{person}{David Vallet} {and} \bibinfo{person}{Hugo
  Zaragoza}.} \bibinfo{year}{2008}\natexlab{}.
\newblock \showarticletitle{Inferring the most important types of a query: a
  semantic approach}. In \bibinfo{booktitle}{{\em Proc. of SIGIR}}.
  \bibinfo{pages}{857--858}.
\newblock


\bibitem[\protect\citeauthoryear{Vercoustre, Pehcevski, and Thom}{Vercoustre
  et~al\mbox{.}}{2008}]%
        {Vercoustre:2008:UWC}
\bibfield{author}{\bibinfo{person}{Anne-Marie Vercoustre},
  \bibinfo{person}{Jovan Pehcevski}, {and} \bibinfo{person}{James~A. Thom}.}
  \bibinfo{year}{2008}\natexlab{}.
\newblock \showarticletitle{Focused Access to XML Documents}.
\newblock \bibinfo{publisher}{Springer}, Chapter Using Wikipedia Categories and
  Links in Entity Ranking, \bibinfo{pages}{321--335}.
\newblock
\showISBNx{978-3-540-85901-7}


\bibitem[\protect\citeauthoryear{Weerkamp, Balog, and Meij}{Weerkamp
  et~al\mbox{.}}{2009}]%
        {Weerkamp:2009:AGL}
\bibfield{author}{\bibinfo{person}{W. Weerkamp}, \bibinfo{person}{K. Balog},
  {and} \bibinfo{person}{E.~J. Meij}.} \bibinfo{year}{2009}\natexlab{}.
\newblock \showarticletitle{A Generative Language Modeling Approach for Ranking
  Entities}. In \bibinfo{booktitle}{{\em Advances in Focused Retrieval, INEX}}.
  \bibinfo{pages}{292--299}.
\newblock


\bibitem[\protect\citeauthoryear{Yosef, Bauer, Spaniol, and Weikum}{Yosef
  et~al\mbox{.}}{2012}]%
        {Yosef:2012:HHT}
\bibfield{author}{\bibinfo{person}{Mohamed~Amir Yosef}, \bibinfo{person}{Sandro
  Bauer}, \bibinfo{person}{Johannes Hoffart~Marc Spaniol}, {and}
  \bibinfo{person}{Gerhard Weikum}.} \bibinfo{year}{2012}\natexlab{}.
\newblock \showarticletitle{{HYENA: Hierarchical Type Classification for Entity
  Names}}. In \bibinfo{booktitle}{{\em Proc. of COLING}}.
  \bibinfo{pages}{1361--1370}.
\newblock


\bibitem[\protect\citeauthoryear{Zhu, Song, and R\"{u}ger}{Zhu
  et~al\mbox{.}}{2008}]%
        {Zhu:2008:IDF}
\bibfield{author}{\bibinfo{person}{Jianhan Zhu}, \bibinfo{person}{Dawei Song},
  {and} \bibinfo{person}{Stefan R\"{u}ger}.} \bibinfo{year}{2008}\natexlab{}.
\newblock \showarticletitle{Focused Access to XML Documents}.
\newblock \bibinfo{publisher}{Springer}, Chapter Integrating Document Features
  for Entity Ranking, \bibinfo{pages}{336--347}.
\newblock
\showISBNx{978-3-540-85901-7}


\end{thebibliography}
